\documentclass[12pt,preprint]{aastex}

\begin{document} 

\title{Kinematic Structure of Molecular Gas around High-mass Star YSO, Papillon Nebula, in N159 East in the Large Magellanic Cloud}
\author{Kazuya Saigo \altaffilmark{1}
Toshikazu Onishi \altaffilmark{1}, 
Omnarayani Nayak \altaffilmark{3}, 
Margaret Meixner \altaffilmark{3,4}, 
Kazuki Tokuda \altaffilmark{1}, 
Ryohei Harada \altaffilmark{1}, 
Yuuki Morioka  \altaffilmark{1}, 
Marta Sewi {\l}o \altaffilmark{10}, 
Remy Indebetouw \altaffilmark{8,9}, 
Kazufumi Torii \altaffilmark{2}, 
Akiko Kawamura \altaffilmark{5},  
Akio Ohama \altaffilmark{2},  
Yusuke Hattori \altaffilmark{2},  
Hiroaki Yamamoto \altaffilmark{2}, 
Kengo Tachihara  \altaffilmark{2}, 
Tetsuhiro Minamidani \altaffilmark{6},
Tsuyoshi Inoue \altaffilmark{7},
Suzanne Madden \altaffilmark{11}, 
Maud Galametz  \altaffilmark{12,13}, 
Vianney Lebouteiller  \altaffilmark{11}, 
C.-H. Rosie Chen \altaffilmark{15},
Norikazu Mizuno  \altaffilmark{5,14}, 
and
Yasuo Fukui \altaffilmark{2} }%

\altaffiltext{1}{Department of Physical Science, Graduate School of Science, Osaka Prefecture University,
1-1 Gakuen-cho, Naka-ku, Sakai, Osaka 599-8531, Japan}
\altaffiltext{2}{Department of Physics, Nagoya University, Chikusa-ku, Nagoya 464-8602, Japan}
\email{saigokz@p.s.osakafu-u.ac.jp}
\altaffiltext{3}{The Johns Hopkins University, Department of Physics and Astronomy, 366 Bloomberg Center, 3400 N. Charles Street, Baltimore, MD 21218, USA}
\altaffiltext{4}{Space Telescope Science Institute, 3700 San Martin Drive, Baltimore, MD 21218, USA}
\altaffiltext{5}{National Astronomical Observatory of Japan, Mitaka, Tokyo 181-8588, Japan}
\altaffiltext{6}{Nobeyama Radio Observatory, 462-2 Nobeyama Minamimaki-mura, Minamisaku-gun,
Nagano 384-1305, Japan}
\altaffiltext{7}{Division of Theoretical Astronomy, National Astronomical Observatory, Japan}

\altaffiltext{8}{Department of Astronomy, University of Virginia, PO Box 400325, Charlottesville, VA,
22904, USA}
\altaffiltext{9}{National Radio Astronomy Observatory, 520 Edgemont Rd, Charlottesville, VA, 22903,
USA} 
\altaffiltext{10}{NASA Goddard Space Flight Center, 8800 Greenbelt Rd, Greenbelt, MD 20771, USA}
\altaffiltext{11}{Laboratoire AIM, CEA, Universite Paris VII, IRFU/Service d'Astrophysique, Bat. 709,
91191 Gif-sur-Yvette, France}
\altaffiltext{12}{Institute of Astronomy, University of Cambridge, Madingley Road, Cambridge CB3
0HA, UK}
\altaffiltext{13}{European Southern Observatory, Karl-Schwarzschild-Str. 2, D-85748 Garching-bei-
M¨unchen, Germany}
\altaffiltext{14}{Department of Astronomy, School of Science, The University of Tokyo, 7-3-1 Hongo,
Bunkyo-ku, Tokyo 133-0033, Japan}
\altaffiltext{15}{Max Planck Institute for Radio Astronomy, Auf dem Huegel 69, Bonn 53121, Germany}

\begin{abstract}

We present the ALMA Band 3 and Band 6 results of $^{12}$CO(2-1), $^{13}$CO(2-1), H30$\alpha$ recombination line, free-free emission around 98~GHz, and the dust thermal emission around 230~GHz toward the N159 East Giant Molecular Cloud (N159E) in the Large Magellanic Cloud (LMC). 
LMC is the nearest active high-mass star forming face-on galaxy at a distance 50~kpc and is the best target for studing high-mass star formation. 
ALMA observations show that N159E is the complex of filamentary clouds with the width and length of $\sim$1~pc and several pc.
The total molecular mass is 0.92 $\times 10^{5}$~M$_{\odot}$ from the $^{13}$CO(2-1) intensity. 
N159E harbors the well-known Papillon Nebula, a compact high-excitation HII region. 
We found that a YSO associated with the Papillon Nebula has the mass of 35M$_{\odot}$ and is located at the intersection of three filamentary clouds. 
It indicates that the formation of the high-mass YSO was induced by the collision of filamentary clouds. 
\citet{Fukui2015} reported a similar kinematic structure toward a YSO in the N159 West region which is another YSO that has the mass larger than 35M$_{\odot}$ in these two regions. 
This suggests that the collision of filamentary clouds is a primary mechanism of high-mass star formation. 
We found a small molecular hole around the YSO in Papillon Nebula with sub-pc scale. 
It is filled by free-free and H30$\alpha$ emission. 
Temperature of the molecular gas around the hole reaches $\sim$ 80~K. 
It indicates that this YSO has just started the distruction of parental molecular cloud. 

\end{abstract}

\keywords{ISM: clouds --- ISM: kinematics and dynamics --- ISM:molecules --- stars: formation}

\maketitle

\section{Introduction}

High-mass stars vastly affect the surrounding environments throughout their lifetime, which results in a rapid dissipation of the natal molecular clouds or, sometimes, in triggering the next-generation star formation.  
They are thus essential ingredients controlling the evolution of molecular clouds in spite of the orders-of-magnitude rareness compared with the lower-mass stars.  
The formation mechanism of high-mass stars has been largely unknown due to the rapid destruction of the parental clouds by stellar feedback.   
In theory, two high-mass star formation models are under study for over 15 years \citep{Zinnecker2007, Tan2014}. 
The first one is the scale-up version of the low-mass star formation. 
High-mass star is formed by the monolithic collapse of a gravitationally unstable massive molecular core (so-called ``Core Accretion Model''). 
The other one is that the forming low-mass star(s) gain the mass to become high-mass star(s) by the competitive mass accretion \citep{Bonnell2004} as reviewed in recent articles (so-called ``Competitive Accretion Model'').
In both models, the mass accretion on the high-mass star must overcome the radiative pressure and the ionization caused by the forming high-mass protostar \citep{Nakano2000, McKee2002, Yorke2002, Krumholz2009, Hosokawa2009}.



The importance of physical interactions between the molecular clouds, cloud-cloud collisions, has been recently addressed by many observational and theoretical studies as a mechanism of the formation of high-mass stars.  
NANTEN2 observations toward the super star clusters (SSCs), Westerlund2, NGC3603 and RCW38, revealed supersonic collisions between two molecular clouds followed by strong shock compression of the molecular gas, leading to the formation of the massive clusters in a short time less than 1~Myrs \citep{Furukawa2009, Ohama2010, Fukui2014, Fukui2015b}. 
Similar results were presented toward single O star in M20 and RCW120  \citep{Torii2011, Torii2015}.  
The theoretical study of collisions of dissimilar clouds was initiated by \citet{Habe1992}, followed by \citet{Anathpindika2010} and \citet{Takahira2014}, indicating that the compression due to the collisions can produce massive molecular clumps.  
The magnetohydrodynamical numerical simulations by \citet{Inoue2013} have shown that the compression excites turbulence and amplifies field strength, leading to increased Jeans mass and a top-heavy dense-core mass function after collision, which explains the high-mass accretion rate required to form a high-mass star.

Recently, ALMA opened a possibility to observe molecular clouds in the external galaxies in a spatially resolved sense with the unprecedented angular resolution and sensitivity.  
Among the galaxies, the Large Magellanic Cloud (LMC) has been considered to be one of the best sites to study high-mass star formation. 
The LMC is close to us at a distance of 50~kpc \citep{Schaefer2008, Grijs2014} and the inclination is almost face-on \citep{Balbinot2015}, which can minimize the effect of the contamination of various objects in the same line of sight.  
The star formation activities in the LMC are quite high as shown by the existence of HII regions including 30 Dor, the largest one in the Local Group. 
Among a sample of nearly 300 Giant Molecular Clouds (GMCs) detected by NANTEN \citep[see also Fukui and Kawamura 2010]{Fukui1999, Fukui2008, Mizuno2001, Yamaguchi2001}, N159 is an HII region with the strongest CO intensity with a total molecular mass of $5.3 \times 10^{5}$M$_{\odot}$.  
N159 is resolved into three prominent clumps; N159 West (N159W) associated with compact HII regions, N159 East (N159E) with developed HII regions, and N159 South (N159S) not associated with HII regions \citep{Mizuno2010}.
The two clumps showing currently active star formation, N159W and N159E, have been observed by ALMA in Cycle 1, and the results of N159W were presented by \citet{Fukui2015}

\citet{Mizuno2010} suggested that the N159W peak represents a pre-star-cluster clump of $\sim 10^{5}$ M$_{\odot}$ from observations in multiple-J rotational transitions of CO at a 10~pc resolution. 
By using ALMA, \citet{Fukui2015} spatially resolved the molecular gas in N159W with a spatial resolution of 0.3~pc $\times$ 0.2~pc in $^{12}$CO(2-1) and $^{13}$CO(2-1), and found that the distribution is highly filamentary with a large clump (N159W-N) toward the N159W peak in the past observations. 
These filamentary clouds show straight or curved distributions with a typical width of 0.5~pc - 1.0~pc and a length of 5~pc - 10~pc. 
They also detected molecular outflows toward two high-mass protostars (YSO-N and YSO-S in their paper) for the first time in the external galaxies, showing that these protostars are quite young with the estimated dynamical time scale of $\sim 10^{4}$ years. 
Impressive in particular are N159W-S, one of the outflows, which is located at the intersection of two spatially overlapping filaments.  
Based on the results, they argued that the two filaments collided with each other and triggered the formation of the high-mass star YSO in N159W-S (YSO-S) in a short time scale of $\sim 10^{5}$ yrs. 

Following \citet{Fukui2015}, it is of high importance to look at the neighboring star forming region, N159E, which is associated with several developed HII regions and molecular clouds are observed as dark lanes in optical and H$\alpha$ images \citep{Mizuno2010} and to explore how star formation is influenced by cloud motion and distribution in N159E, where the ``Papillon Nebula'' is an intriguing unique compact HII region \citep{Heydari1999, Meynadier2004}. 
In addition, Chandra observations revealed a large X-ray bubble toward the largest HII region, which is considered to represent supernova remnants \citep{Seward 2010} suggesting that the molecular gas may be affected by the past explosive events.  

In this paper, we present the results of ALMA observations of molecular clouds in N159E.  
The focuses will be on the filamentary distribution of the molecular clouds and the formation of the high-mass star "Papillion Nebula", and a comparison with N159W will broaden and deepen our understanding of the evolution of GMCs in the LMC. We describe observations in Section 2 and results including YSO properties in Section 3. We then discuss on the triggering of high-mass star formation in Section 4 and conclude the paper in Section 5.


\section{Observations}

We carried out ALMA Cycle 1 Band 3 (86-116~GHz) and Band 6 (211-275~GHz) observations toward N159E both with the main array 12m antennas and the Atacama Compact Array (ACA) 7m antennas.
This observation is based on ALMA High Priority Project 2012.1.00554.S (PI:Yasuo Fukui). 
This project consists of mapping observations of the N159E and N159W molecular clumps in LMC using the same observational setting (spectral setting, spatial resolution and sensitivity).  
Results of N159W which is the counterpart result of this paper are shown by \citet{Fukui2015}. 
For N159E, we have observed the main body of the molecular clump in N159E region by using the mosaic mapping of $60'' \times 75''$ rectangular region around the central position of ($\alpha_{\rm J2000.0}$, $\delta_{\rm J2000.0}$) = (05$^{\rm h}$40$^{\rm m}$08$^{\rm s}$.0429, -69$^{\rm d}$44$^{\rm m}$39$^{\rm s}$.515). 
Our target molecular lines were $^{13}$CO(J=1-0), C$^{18}$O(J=1-0), CS(J=2-1), $^{12}$CO(J=2-1), $^{13}$CO(J=2-1) and C$^{18}$O(J=2-1) with the frequency resolution of 15.3~kHz ($\times$ 3840~channels). 
We also observed the continuum emission with the wide bandwidth of 1875.0~MHz (488.3~kHz 3840~channels). 
The radio recombination lines of H30$\alpha$ and H40$\alpha$ were included in the wide bandwidth observations.
The ALMA Band 6 observations were carried out on December 2 2013 and March 7 2014 with total on-source time of 25~minutes. 
It used 27 antennas and the projected baseline length of the 12m array range is from 13~m to 392~m. 
The ALMA Band 3 observation was carried out on December 3 2013 with total on-source time of 5~minutes. 
It used 26~antennas and the projected baseline length of the 12m array range is from 15~m to 452~m. 
In the both Bands, projected baselines of ACA observation covers from 8~m to 36~m. 

In calibration processes, we made additional flagging of a few antennas that have too low gain or show large amplitude dispersion in time. 
The bandpass calibration of the 12m array was carried out by using 4 quasars (J1058+0133 and J0538-4405 in Band 6, J0006-7516 and J0334-4008 in Band 3). 
The complex gain calibration was carried out by using 3 quasars (J0635-7516 and J0601-7036 in Band 6, J0635-7516 in Band 3). 
These quasars have a sufficient flux density for the calibration. 
Flux calibration was carried out by using 3 solar system objects (Pallas and Ganymede in Band 6, Uranus in Bnad 3). 
For the flux calibration, we used the Butler-JPL-Horizons 2012 model \\
({\small https://science.nrao.edu/facilities/alma/aboutALMA/Technology/ALMA Memo Series/alma594/abs594}).
The calibration and data reduction were made using the Common Astronomy Software Application (CASA) package (http://casa.nrao.edu) and visibility imaged. 
We used the natural weighting for both the Band 6 and Band 3 data, providing synthesized beam sizes of 1.21'' $\times$ 0.84'' (0.3~pc $\times$ 0.2~pc at 50~kpc) and 2.68'' $\times$ 1.66'' (0.7 $\times$ 0.4~pc at 50~kpc), respectively. 
The rms noises fluctuations of molecular lines with the velocity resolution of 0.2~km s$^{-1}$ are 46~mJy beam$^{-1}$ in Band 6 and 26~mJy beam$^{-1}$ in Band 3, in emission-free channels.

\section{Results}

\subsection{Filamentary distribution of molecular gas toward N159E in $^{12}$CO(2-1) and $^{13}$CO(2-1)} \label{subsec:N159Eall}

Figure \ref{N159E_mom0} shows the velocity-integrated intensity image of N159E in $^{12}$CO(2-1) and $^{13}$CO(2-1) obtained with the 12m array of ALMA.  
The total fluxes of $^{12}$CO(2-1) emission with the 12m array and with the 7m array are almost the same with the difference less than 10~\%, and thus we use only the 12m array data here because no significant resolved-out emission is expected due to the interferometric observations.  
Figure \ref{N159E_mom0}(a) shows that molecular clouds are distributed throughout N159E and the prominent feature is three belt-like structures extending from the northeast to the southwest with length of 10~pc - 20~pc. 
These cloud complexes are resolved into smaller scale filamentary/clumpy features. 
Many of them are also traced by the denser gas tracer, $^{13}$CO(2-1) emission (see Figure\ref{N159E_mom0}(b)).

We estimated the mass and column density of the N159E cloud using two different methods. 
The total mass is estimated to be $ (1.3 \pm 0.4) \times 10^{5}$~M$_{\odot}$ from the $^{12}$CO(2-1) intensities by the same method as \citet{Fukui2015}.  
In this estimation, we assumed the typical $^{12}$CO(2-1)/$^{12}$CO(1-0) ratio toward HII regions of 0.85 (the ratio in the Orion-KL region of \citet{Nishimura2015}) and a conversion factor from the $^{12}$CO(1-0) intensity to the column density of X(CO) = $(7 \pm 2) \times 10^{20}$~cm$^{-2}$ \citep{Fukui2008}. 
The detection limit in column density is $\sim 2.0 \times 10^{21} $~cm$^{-2}$ and the maximum column density is $\sim 2.0 \times 10^{23}$~cm$^{-2}$. 
The mass estimation from the $^{12}$CO intensity may be inaccurate in the optically thick region of the  high column density or the heated cloud by the high-mass star (see section \ref{subsec:Papillon}). 
So, we estimated the mass also from $^{13}$CO(2-1) intensity by assuming the local thermodynamical equilibrium (LTE) and the $^{13}$CO/H$_{2}$ abundance ratio of $3.2 \times 10^{-7}$ \citep{Fujii2014}. 
\citet{Fujii2014} assumed a uniform excitation temperature of 20~K, because the large beam size diluted the line intensities, hampering the estimation of absolute peak temperatures; the present ALMA observations spatially resolved the molecular distribution.
The excitation temperatures were thus calculated from the peak $^{12}$CO(2-1) temperature by assuming that the line is optically thick, and were assumed to be 20~K where the calculated excitation temperatures were below 20~K (e.g., Nishimura et al. 2015). 
The total molecular mass of N159E in the present map is then calculated to be $0.92 \times 10^{5}$~M$_{\odot}$ from the $^{13}$CO(2-1) intensity. 
It is consistent with the mass derived from $^{12}$CO(2-1).

White cross and dot symbols denote the position of four high-mass YSO candidates which are cataloged in \citet{Gruendl2009}, \citet{Chen2010} and \citet{Carlson2012}.
The strongest CO intensity clump of $^{12}$CO(2-1) and $^{13}$CO(2-1) is associated with the high-mass YSO, J054009.40-694437.6 that is pointed as a large cross symbol. 
This YSO is also associated with a compact HII blob called the Papillon Nebula \citep{Heydari1999, Meynadier2004} (hereafter, we call it ``Papillon Nebula YSO'' in this paper). 
The high-mass YSO candidate, J054009.49-694453.5, is located at a local peak of the molecular cloud.
Two white dot symbols denote the position of newly identified high-mass YSO candidates by \citet{Carlson2012}. 
They are not associated clearly with the dense molecular cloud but still possess remnant circumstellar material (see the details in Section \ref{subsec:YSO}).

Figure \ref{N159_Halpha} shows H$\alpha$ and CO intensity distributions toward the N159 cloud. 
The color image shows the H$\alpha$ intensity distribution based on observations made with ESO telescopes at the La Silla Observatory programme ID 07.C-0888; processed and released by the ESO VOS/ADP group with the logarithmic scale (see Mizuno et al. 2010).
Gray contour shows the integrated intensity of $^{12}$CO(J=3-2) with ASTE 10m telescope \citep{Minamidani2008}.  
The N159E and N159W cloud are located at the eastern and western edges of the N159 HII bubble. 
Black contours show the integrated intensity of $^{12}$CO(J=2-1) toward N159E and N159W with ALMA Band 6 (see Fukui et al. 2015 for N159W) with the contour levels of 10 and 30 sigma (1~sigma = 0.3~Jy/Beam kms$^{-1}$). 
ALMA mapping area of this observation is shown by the black dotted contours. 
The distribution of molecular cloud is well coincident with the dark lane in the optical H$\alpha$ image except for the southern part of N159W. 
Most of high-mass YSO candidates, especially YSO candidates with the mass of more than 30~M$_{\odot}$ (large cross smbols) are clearly associated with the molecular cloud.
These results suggest that the main body of the molecular clouds is located in front of the N159 HII bubble and the destruction of the molecular cloud caused by the high-mass YSOs has not been progressed. 
Both region are in the early stage of high-mass star formation and are in the similar environment.

Figure \ref{N159E_channel} shows the velocity channel maps of the N159E cloud in $^{12}$CO(2-1) and $^{13}$CO(2-1). 
It shows that the N159E cloud consists of a large number of filamentary clouds with different velocities. 
The typical width and projection length of the filaments are $\sim$1~pc and 5~pc - 10~pc, respectively. 
Many of the filaments are aligned roughly in the northeast -- southwest direction.
Some of the filaments that are associated with high-mass YSO candidates have north-south direction. 
Especially, filaments associated with the Papillon Nebula have a unique kinematic structure.
We show the kinematic structure of the filaments associated with Papillon Nebula YSO in detail in section 3.3.


\subsection{YSO characteristics}\label{subsec:YSO}

In recent years, YSO candidates have been photometrically and spectroscopically identified in the LMC (Whitney et al. 2008, Gruendl \& Chu 2009, Seale et al. 2009, Meixner et al. 2013, Carlson et al. 2012, Seale et al. 2014). 
However it still remains unclear how the formation of YSOs is associated with the surrounding molecular gas. 
ALMA Cycle 1 observations help us in relating the properties of the YSOs to the properties of the gas in which they reside. 
Figure \ref{N159_Halpha} shows that several high-mass YSOs are currently forming in the N159 region (Gruendl \& Chu 2009, Chen et al. 2010, Carlson et al. 2012, Seale et al. 2014, Fukui et al. 2015). 
There are a total of 4 YSOs in our ALMA footprint of N159E. 
In order to derive the characteristics of the YSOs, we use the 2D radiative transfer YSO model grid of Robitaille et al. (2007) and the spectral energy distribution (SED) fitter tool to determine physical parameters of YSOs from available photometry data with Spitzer and Herschel fluxes 
using the 2MASS near-IR (JHKs), Spitzer/SAGE (3.6-24), and Herschel/HERITAGE (100-500 um) photometry 
included in our modeling.

Table \ref{t1} shows the physical properties of YSOs in the N159E. 
Mass, Luminosity and age are the best fit results of SED fitting. 
The first two objects in Table \ref{t1}, also shown as crosses in Figure \ref{N159E_mom0}, are both high-mass YSOs with a significant amount of CO gas surrounding them.  
The fit of the SEDs indicates that both of the sources are Stage 0/I YSOs and the Papillon Nebula YSO had a nearly pole-on orientation. 
The mass and luminosity are estimated to be 35 M~$_{\odot}$ and $1.8 \times 10^{5}$~L$_{\odot}$  for the Papillon Nebula YSO, and 18~M$_{\odot}$  and $4.5 \times 10^{4}$~L$_{\odot}$ for J054009.49-694453.5.

\citet{Chen2010} estimated that the mass and luminosity of the Papillon Nebula YSO is 20~M$_{\odot}$ and $5.3 \times 10^{4}$~L$_{\odot}$, respectively. 
However, \citet{Chen2010} also concluded that the Papillon Nebula YSO is multiple sources since single-YSO model fails to reproduce the SED simultaneously using the flux between optical and mid-IR emission. 
This best fit mode is biased on the flux of shorter wavelength ($\lambda < 8~\mu m$) due to the lack of data and it cannot reproduce the flux of longer wavelength (see Fugure 7 of Chen et al. 2010) . 
Note that they also show that the mid-IR part of the SED can be reproduced by more massive models with $M_{star} \sim 41~M_{\odot}$ and it is consistent with the stellar mass derived from radio observations by \citet{Indebetouw2004}.
Our fittings result in higher mass and luminosity due to the addition of Herschel PACS and SPIRE photometry (Seale et al. 2014). 
This may be a more plausible estimate of the YSO mass because the PACS and SPIRE photometry completes the SED and thereby our fit captures the emissions from all the surrounding dust. 
Unlike N159W-N YSO and N159W-S YSO (Fukui et al. 2015a, YSO-N and YSO-S in their paper), neither the Papillon Nebula YSO nor J054009.49-694453.5 have $^{12}$CO(2-1) outflow wing detections. 
The other two YSOs in N159E, confirmed by \citet{Carlson2012} and marked by white dots in Figure \ref{N159E_mom0}, are less massive YSOs with not as much $^{12}$CO(2-1) or $^{13}$CO(2-1) gas detected near them.

The Papillon Nebula YSO is associated with radio recombination line detections. 
We determine the electron temperature of the gas using the H30$\alpha$ emission \citep{Sewilo2011}.
We then use the continuum emission and equations provided by Mezger \& Henderson (1967) to determine the electron density, ionized mass, emission measure, ionization parameter, and Lyman continuum flux (all of which are dependent on the electron temperature). 
We list physical properties in Table \ref{t2}. 
Here, the N159W-N YSO is the another high-mass YSO that is associated with H30$\alpha$ detection in our ALMA observation toward N159 (see figure \ref{N159_Halpha}) and it is a good high-mass YSO for comparison. 
Both the Papillon Nebula YSO and the N159W-N YSO have similar ionizing properties and are at least an O3 star (Smith et al. 2002). 
However the Papillon Nebula YSO has an emission measure value half of that of the N159W-N YSO. 
The emission measure is proportional to the optical depth. 
Our calculations using  H30$\alpha$ and the continuum emission are consistent with the Robitaille SED fitting results. 
There is less circumstellar material along the line-of-sight (i.e. pole-on orientation), as for the case of the Papillon Nebula YSO compared with to N159W-N YSO. 
Therefore we find a smaller emission measure with the Papillon Nebula YSO which is indicative of a lower optical depth.

Meynadier et al. (2004) use evolutionary track models (Lejeune \& Schaerer 2001) to determine that the location of the Papillon Nebula YSO in their color-magnitude diagram best fits a 3 Myrs isochrone and the mass of the Papillon Nebula YSO is 50~M$_{\odot}$. 
Our calculations are similar to the previous calculations of mass, photon flux, and spectral type. 
The differences may be mainly due to uncertainties in extinction, measured fluxes, and theoretical evolutional tracks. 
We use mid- to far-IR photometry for SED fitting and the continuum observation to calculate the spectral type, therefore we are assuming that none of ionizing photons are escaping. 
However this assumption does not always hold. 
We observe a CO-hole around the Papillon Nebula YSO in the present observations (see figure \ref{N159E_Papillon}a and section 3.3), but this is not the case for N159W-N YSO. 
Papillon Nebula YSO is surrounded by much lower dense gas compared to N159W-N YSO. 
More photons could be escaping the Papillon Nebula YSO without ionization and our calculations may underestimate the Lyman continuum flux and the stellar mass. 
This scenario is further supported by Meynadier et al. (2004) who found the Papillon Nebula YSO to be not very embedded (Av  $\sim$ 7~mag) and to be situated on the nearer side of the cloud to us since it is visible in the optical.

There are 9 YSOs in N159W \citep{Fukui2015}, one of which only has Herschel PACS and SPIRE photometry and therefore not listed in Table \ref{t3}. 
J053943.75-694540.2, J053940.48-694517.3, and J053940.78-694632.0  are all less than 15~M$_{\odot}$. 
These three YSOs have a dearth of  $^{12}$CO(2-1) or $^{13}$CO(2-1) gas near them, similar to the two low-mass YSOs in N159E. 
Perhaps lower mass YSOs can form in regions that lack $^{12}$CO(2-1) or $^{13}$CO(2-1) however they are probably not formed via filamentary collisions. 
There are 3 YSOs in N159W and 1 YSO in N159E that have masses above 30~M$_{\odot}$: N159W-N YSO, N159W-S YSO, J053937.53-694609.8, and the Papillon Nebula YSO. N159W-N YSO and N159W-S YSO are $\sim 10^{4}$~years old and have outflow associated with them \citep{Fukui2015}.  
J053937.53-694609.8 and the Papillon Nebula YSO have a CO-hole around them, are both $\sim 10^{5}$~years old, and have no outflow wings detected from the CO data. 
It is plausible to assume that at one point in time both J053937.53-694609.8 and the Papillon Nebula YSO had outflows, they are now older and are accreting less material with have dissipated CO gas around them. 
Analyses of ALMA $^{12}$CO(2-1) and $^{13}$CO(2-1) gas associated with YSOs in other regions of the LMC will shed more light on the difference between higher mass and lower mass star formation and evolution.

\subsection{Gas properties toward the Papillon Nebula}\label{subsec:Papillon}

Figure \ref{N159E_Papillon}(a) shows the H$\alpha$ image of WFPC2 HST around the Papillon HII blobs (the white rectangle region in figure \ref{N159E_mom0}). 
The H$\alpha$ observations revealed the morphology of a ``papillon'', a butterfly-shaped ionized nebula with the wings separated by $\sim$ 2.3" (0.6~pc) \citep{Heydari1999}. 
White contours show the velocity integrated map of $^{12}$CO(2-1). 
The distribution of molecular gas toward the Papillon Nebula shows a CO-hole with a diameter of $\sim 0.6$~pc.  
Blue contours show the radio recombination line, H30$\alpha$, detected by our ALMA Band 6 observation. 
Distribution of H30$\alpha$ is tracing the ionized gas, which almost fills the molecular hole. 
Green contours show the continuum emission at 98~GHz of ALMA Band 3. 
The 98~GHz continuum emission seems to be dominated by free-free emission from ionized gas since it is several times stronger than the expected thermal dust emission by 231~GHz continuum emission. 
In this paper, we define the central position of the Papillon Nebula is the central position of 98~GHz continuum emission of ALMA Band 3 (magenta cross) because it has the highest spatial position accuracy among optically thin emission from ionized gas. 
Figure \ref{N159E_Papillon}(b) shows the velocity integrated intensity image of $^{13}$CO(2-1), showing the distribution of dense gas. 
The red contour shows the distribution of 231~GHz thermal dust continuum emission observed by ALMA Band 6. 
The dust emission is located at the edge of the molecular hole.  
This indicates that the cloud surface is heated by the strong radiation from high-mass YSO in the Papillon Nebula.

Figure \ref{N159E_Papillon_Slice} shows the antenna temperature distribution along slices with three different position angles to demonstrate distribution of Papillon Nebula and the interacting molecular cloud. 
The distribution of $^{12}$CO(2-1) has a maximum at 2 - 3'' (0.5 - 0.75~pc) from the center of Papillon Nebula. 
The maximum antenna temperatures on the slices of PA = 23~$^{\circ}$, -37$^{\circ}$ and -97$^{\circ}$ reach 76~K, 79~K and 65~K, respectively.  
These temperature peaks are located at the molecular cloud edge toward the Papillon Nebula, suggesting that the molecular gas is heated by the UV from the Papillon Nebula YSO.
Molecular line intensity is steeply dropped toward the Papillon and forms a molecular hole. 
The molecular hole is filled by 98~GHz continuum emission (red solid lines). 
Except for direction of P.A. = +23~$^{\circ}$, 231~GHz thermal dust emission (red dashedlines) has a peak at the surface of molecular cloud. 

This is in contrast to the situation toward two YSOs, N159W-N YSO and N159W-S YSO, in N159W, which are located at the peak of the CO clouds.  
This fact is also consistent with that the Papillon Nebula is in a later evolutionary stage than the N159W YSOs.  
The lack of the outflowing gas also supports the evolved nature of the source, indicating that the mass accretion onto the YSO in the Papillon Nebula has halted.

We don't see significant absorption in the H$\alpha$ emission image toward the region where the CO is detected.  
The column density of the molecular gas toward the H$\alpha$ emission is as high as $2 \times 10^{23}$~cm$^{-2}$, implying that the Papillon is located in front of the molecular clouds facing to us.  
This fact is consistent with the discussion by \citet{Meynadier2004} as shown in Section \ref{subsec:YSO}. 

The high antenna temperature of $^{12}$CO(2-1) toward the interface region can be explained that the heated surface of the molecular cloud is on the near side.  
The mass of the ionized dense gas is estimated to be 65~M$_{\odot}$ from the H30$\alpha$ observations (Table \ref{t2}).  
The total flux of 231GHz continuum emission around Papillon Nebula within the radius of 1pc is $\sim$ 90~mJy. 
The mass of the dust cloud is then calculated to be 5340~$\displaystyle(\frac{T_{\rm dust}}{80 \rm K})^{-1}$~M$_{\odot}$ by assuming that the continuum is only from the dust thermal emission.
Here, we assumed the absorption coefficient per unit dust mass at 1.2~mm and the dust-to-gas mass ratio is 0.77~cm$^2$ g$^{-1}$ and $3.5 \times10^{-3}$, respectively \citep{Herrera2013, Fukui2015}.  
Figure \ref{N159E_Papillon}(b) shows that about half of the dust emission comes from the CO hole, indicating that the gas mass without CO emission is as large as a few thousand ~M$_{\odot}$ inside the CO hole.


\subsection{Filaments toward Papillon Nebula}\label{subsec:Filaments}

  Previous observations of molecular clouds in high-mass star forming regions showed that multiple velocity components were often observed toward high-mass YSOs/clusters, and this fact is considered to be a possible consequence of the past cloud-cloud collisions \citep{Furukawa2009, Ohama2010, Torii2011, Torii2015, Fukui2015, Fukui2015b}. 
Also in case of N159E, the complex velocity structure is seen toward the molecular cloud associated with the Papillon.  
We then focus on this region in order to investigate the origin of the high-mass YSOs. 

Figures \ref{N159E_Papillon_Filaments}(a) and 5(b) show kinematic structures of the associated clouds with Papillon Nebula. 
Although the velocity integrated intensity image in Figure \ref{N159E_Papillon}(b) shows that the cloud associated with the Papillon Nebula looks like a single clump, the velocity structure suggests that it consists of differernt celosity component.  
The velocity distribution in $^{12}$CO(2-1) shown in Figure \ref{N159E_Papillon_Filaments}(a) indicates that the velocity of the molecular gas varies by $\sim$ 10km s$^{-1}$ in the cloud complex. 
The velocity channel maps in Figure \ref{N159E_channel} indicate that the complex can be divided into a few filaments with different velocities.
Figure \ref{N159E_Papillon_Filaments}(b) shows the three velocity components; blue (-8.0~km s$^{-1}$ -- -1.6~km s$^{-1}$), green (-1.4~km s$^{-1}$ -- +1.0~km s$^{-1}$), and red (+1.2~km s$^{-1}$ -- +7.8~km s$^{-1}$). 
Each velocity component has filamentary structures, and they apparently merges around the region where the hole is located.  
This morphology and the size scale are similar to the N159W-S filament that shows high-mass star formation induced by two filaments collision (see figure \ref{N159E_Papillon_Filaments}(c)) although the situation is more complex than in the N159E filaments.

Figure \ref{N159E_Papillon_PV} shows the PV diagrams along each filament.  
The slice regions of the PV diagrams are shown as dashed magenta rectangles in Figure \ref{N159E_Papillon_Filaments}(b).
Velocity is the relative velocity against the system velocity of N159E and the position denotes the offset positions from the center of Papillon Nebula (magenta cross symbol in Figure \ref{N159E_Papillon_Filaments}(b)). 
Each filament clearly has two regions with different line width. 
The line width measured as Full Width at Half Maximum (FWHM) of each filament is 4 - 6~km~s$^{-1}$ in a place near the Papillon and is 2 - 3~km~s$^{-1}$ in a place far from the Papillon. 
Within the 3'' ($\sim$ 0.75~pc) from the Papillon, the centroid velocity of all filamentary clouds are merged into the velocity of blue filament ($\sim$ -2~km~s$^{-1}$) .

\section{Discussion: Filamentary collision as a mechanism of formation of high-mass stars}

The present work has shown that the molecular gas in the N159E region is remarkably filamentary. 
The typical width and projection length of the filaments are $\sim$1~pc and 5~pc - 10~pc, and the filamentary nature is similar to the N159W region \citep{Fukui2015}, suggesting that the filamentary distribution is common in GMCs. 
The most outstanding young star in N159E is the Papillon Nebula. 
We shall hereafter focus on Papillon and discuss star formation properties which are characteristic to the filamentary cloud distribution in N159.
Papillon is located toward an intersection of the three filamentary clouds as shown in Figures \ref{N159E_Papillon_Filaments}(a) and \ref{N159E_Papillon_Filaments}(b).
We find three major differences between Papillon and N159W-S; 
(1) Papillon Nabula YSO is associated with the compact HII region, whereas N159W-S YSO has no detectable HII region, 
(2) N159W-S YSO is associated with very young protostellar outflow although the Papillon YSO has no outflow, and 
(3) Papillon Nebula YSO is associated with the third filament that is nearly orthogonal to the other two filaments, while N159W-S YSO is associated with two filaments. 
The stellar masses of the N159W-S YSO and Papillon Nebula YSO are $\sim$40~M$_{\odot}$ and $\sim$30~M$_{\odot}$, respectively, and both of them are O stars capable to form an HII region. 
These facts are consistent with that the Papillon is in an evolutionary stage later than N159W-S, when outflow is halted and an HII region is developed.
\citet{Fukui2015} suggested that the V-shaped filaments collided with each other to trigger the O star formation in N159W-S. 
It is natural that a similar filamentary collision, possibly a triple collision, took place in Papillon. 
In particular, the two filaments, Green and Red, show remarkably similar velocity distribution to the two filaments in N159W-S, where the intersection between the two filaments remains unionized (Figure 3 of Fukui et al. 2015). 
The molecular filaments in Papillon are not extended toward the O star due to the ionization. 
The third Blue filament may be colliding also with the other two. 
The collisional interaction is consistent with the line broadening near the Papillon Nebula in the PV diagrams in Figure \ref{N159E_Papillon_PV} as discussed in case of N159W-S; cloud-cloud collision excites isotropic turbulence whose velocity span is similar to the true relative velocity \citep{Inoue2013}. 
The velocity span in the broadened region is as large as ~9 km s$^{-1}$, suggesting the relative velocity of the collision is about 9 km s$^{-1}$. 
We see signs of the line broadening even at an offset about 11~arcsec and 7~arcsec from the O star in Red and Green where the HII region is not extended (Figure \ref{N159E_Papillon_PV}). 
The broadening, thus, seems not to be due to the dynamical effect of the HII region. 
We in addition note that the Red filament is associated with weak emission filling the area between the Red and Green filaments. 
Similar weak emission is also seen in case of N159W-S (Figure \ref{N159E_Papillon_Filaments}). 
We suggest that these weak emissions may represent the low-density envelopes of the filaments, which are heated-up mildly by the collisional shock interaction.
The above scenario suggests the time scale of the collision is estimated to be 2pc divided by 9~km s$^{-1}$, i.e., $2 \times 10^{5}$~yrs., which is similar to the time scales of the collision and the stellar age of Papillon Nebula YSO.  
We suggest that formation of the O star, i.e., mass-accretion onto the protostellar core, was initiated just after the filamentary collision.  
The O star associated with the Papillon Nebula is more evolved than N159W-S whose outflow has a dynamical timescale of $10^{4}$ yrs.
For a stellar mass of 30~M$_{\odot}$ and the timescale $2 \times 10^{5}$~yrs, the mass accretion rate is estimated to be $\sim 10^{-4}$~M$_{\odot}$ yr$^{-1}$ . 
This mass accretion rate satisfies the requirement to overcome the radiation pressure \citep{Wolfire1987}.

The small velocity difference between the two filaments suggests that the relative motion makes an angle to the line of sight rather close to 90$^{\circ}$ as in case of N159W-S. 
This in turn indicates that the angle between the two filaments can be significantly affected by the projection and may be much larger than the projected one, $\sim$20$^{\circ}$. 
The similar filamentary collision both in N159W and N159E suggests that such a collision may be frequent in N159 and in the LMC, being possibly a major mode of O star formation. 
If we assume tentatively a collision between two filaments of 10~pc length, which move in a direction vertical to their elongation at a relative velocity of 10 km s$^{-1}$ with an angle of 90 degrees projected on a plane normal to the relative motion, the volume where filamentary collision takes place can be large (10pc)$^3$ in 1~Myrs for a GMC size of 20~pc, in the order of 10~\% of the total volume of the GMC (20~pc)$^3$. 
Given the typical evolutionary timescale of a GMC 10~Myrs \citep{Fukui2010}, such collision should happen fairly frequently as is consistent with the observations.
It is intriguing that Papillon and N159W-S shows similar morphology of the colliding V-shaped filaments (Figure \ref{N159E_Papillon_Filaments}) in spite of the separation of 100~pc in projection. 
We suggest that the whole N159 region including N159W and N159E may be subject to large-scale acceleration or compression driven over a few 100~pc scale supershell(s). 
We speculate that such a compression may help to organize magnetic field in a common direction both in N159W and N159E.

\section{Conclusions}\label{sec:conclusions}

We observed the N159E giant molecular cloud in the LMC with the ALMA Band 3 and Band 6. 
In this paper, we showed the results of $^{12}$CO(2-1), $^{13}$CO(2-1), H30$\alpha$ hydrogen recombination line, free-free emission around 98~GHz, and dust thermal emission around 230~GHz. 
Conclusions of the present work are given as follows.

\begin{figure}
 \begin{center}
  \includegraphics[width=18cm]{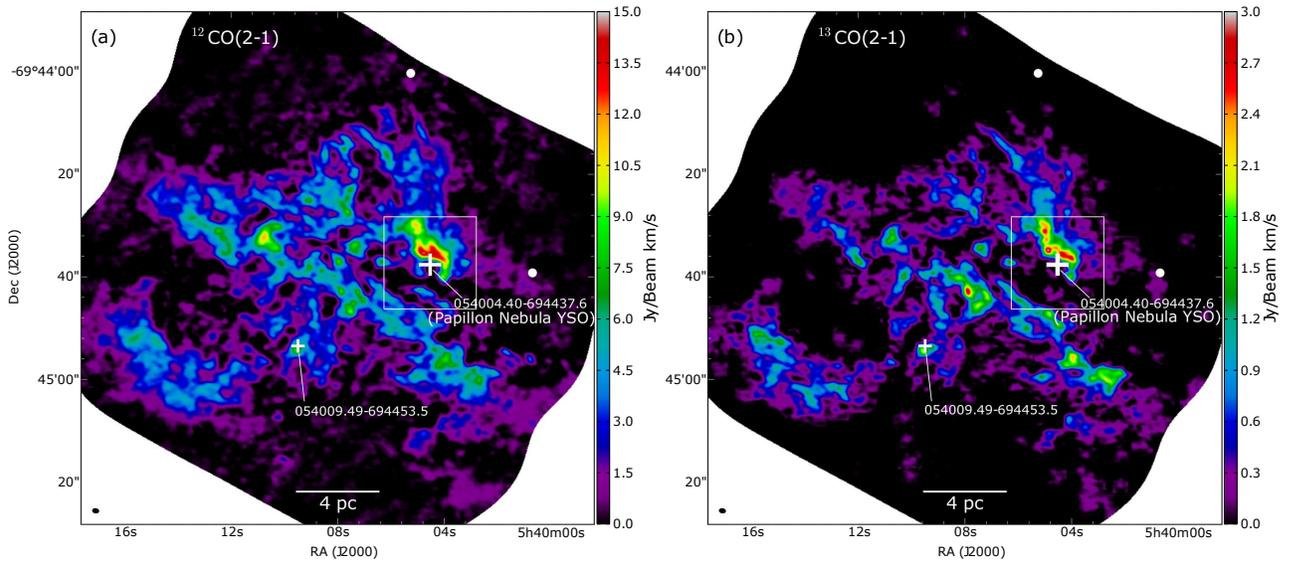}
   \end{center}
 \caption{Integrated intensity map of (a) $^{12}$CO(J=2-1) and (b) $^{13}$CO(J=2-1) toward the N159E region observed with the ALMA 12m array. 
Ellipse at the bottom-left corner in both panel is the synthesized beam (1.21'' $\times$ 0.84'' P.A. =~82$^{\circ}$). 
Large and small white cross symbols denote the positions of a high-mass YSO candidate of $ M_{star} > \sim$~30~M$_{\odot}$ and 8~M$_{\odot} < M_{star} < $ 30~M$_{\odot}$, respectively (see table \ref{t1}).  
White dot symbols denote newly identified high-mass YSO candidates by \citet{Carlson2012}.
}
\label{N159E_mom0}
\end{figure}

\begin{figure}
 \begin{center}
  \includegraphics[width=18cm]{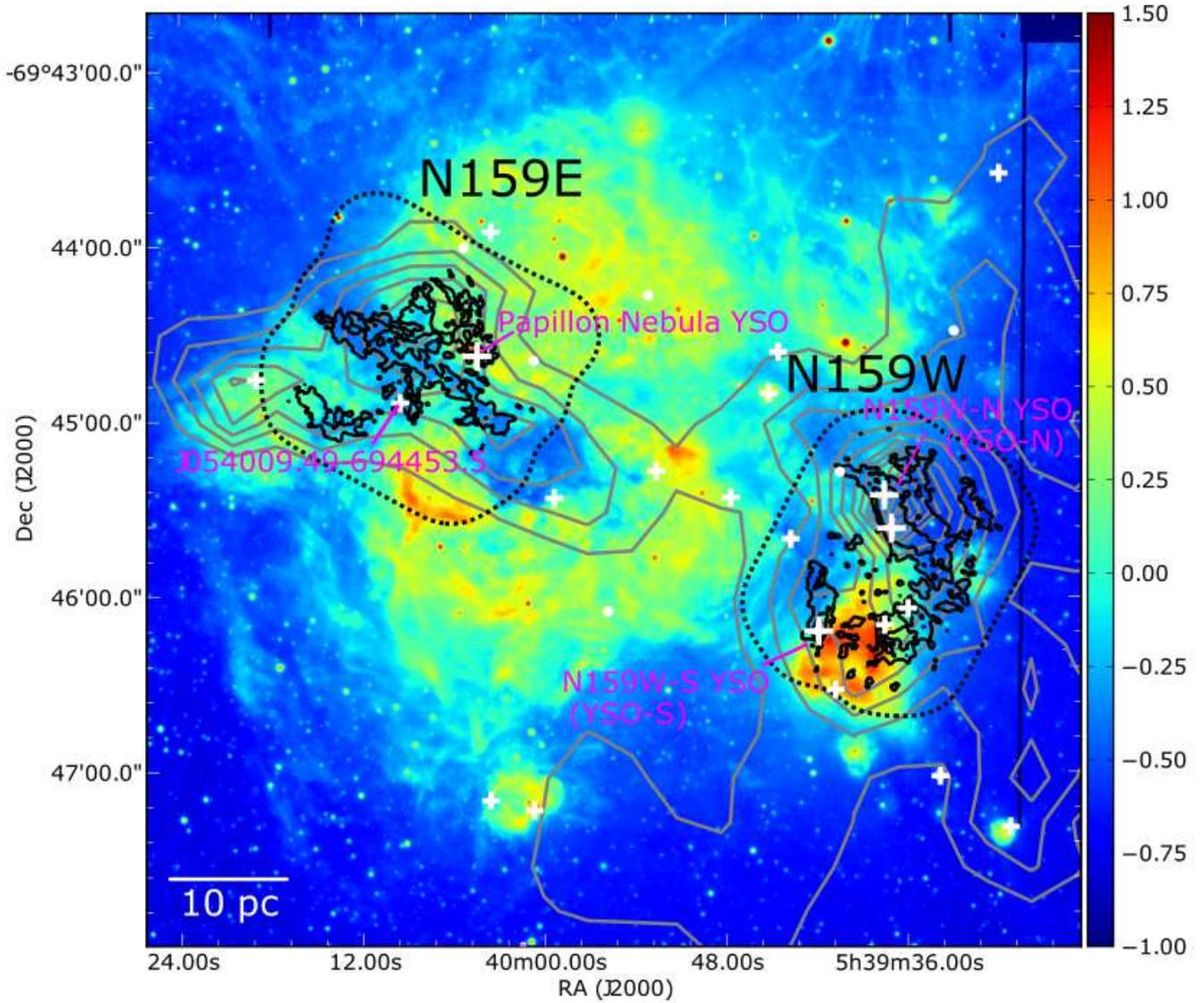}
   \end{center}
 \caption{H$\alpha$ image and CO integrated intensity toward the N159. 
Black and gray contour show the integrated intensity map of $^{12}$CO(J=2-1) with ALMA Band6 and $^{12}$CO(J=3-2) with ASTE 10m telescope, respectively.  
Black dotted contour shows the mapping area of our ALMA Band6 observation. 
}
\label{N159_Halpha}
\end{figure}

\begin{figure}
 \begin{center}
  \includegraphics[width=18cm]{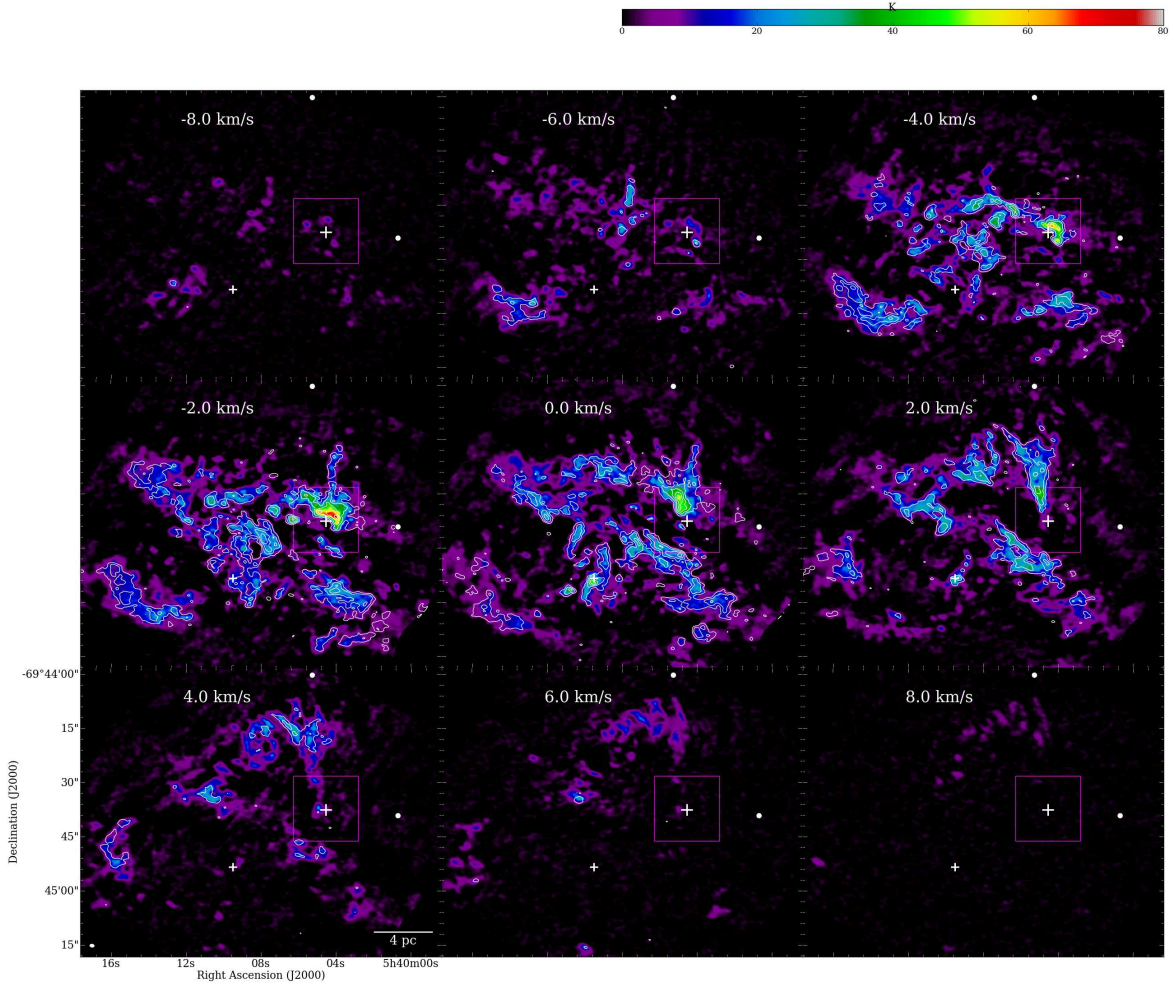}
 \end{center}
 \caption{
Velocity channel maps of N159E in $^{12}$CO(2-1) and $^{13}$CO(2-1). 
The velocity width for each map is 2.0km s$^{-1}$. 
Velocity in each panel denote the relative velocity against the system velocity of N159E, $v - v_{\rm sys}$. 
In this paper, we use the system velocity of N159E of 234.2~km~s$^{-1}$ that is the peak velocity of $^{12}$CO(2-1) of N159E clump.  
White contour show velocity channel maps of 13CO(J=2-1). Contour levels are 5, 15, 30, 45, 60$\sigma$. 
The noise levels are $1 \sigma = 0.33$K and 0.347K in $^{12}$CO(2-1) and $^{13}$CO(2-1), respectively.  
White symbols show the position of high-mass YSO candidates.}
\label{N159E_channel}
\end{figure}

\begin{figure}
 \begin{center}
  \includegraphics[width=18cm]{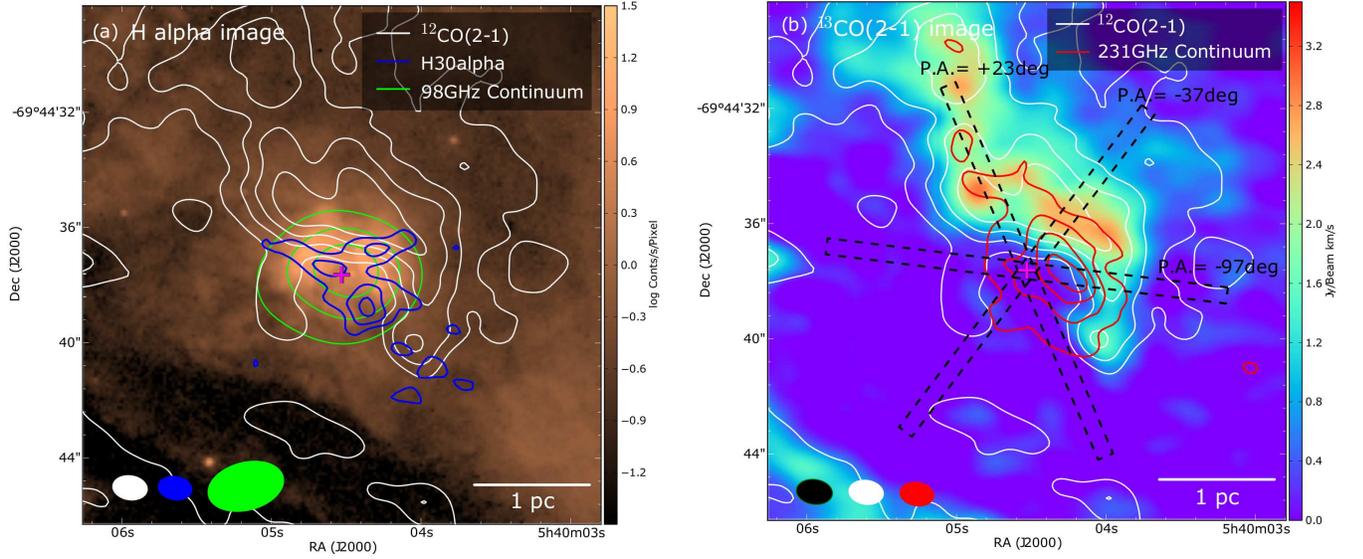}
 \end{center}
 \caption{(a) Ha image of the Papillon HII region by WFPC2 of HST. 
The field size is 18arcsec x 18arcsec (4.5p x 4.5pc). 
White contour show the integrated intensity of 12CO(2-1). 
Contour levels start from 10sigma in steps of 10 (1 sigma = 0.3 Jy/Beam kms$^{-1}$). 
Green and blue contours show 98GHz continuum emission of ALMA Band3 and the integrated intensity of H30$\alpha$ of ALMA Band6, respectively. 
Red cross shows the central position of 98GHz continuum emission. 
(b) Integrated intensity map of $^{13}$CO(J=2-1). 
Red contour show 231GHz continuum emission of ALMA Band6. 
Black rectangular denotes slice paths of Papillon HII region and the surrounding molecular cloud with three different position angles (see Figure \ref{N159E_Papillon_Slice}). }
\label{N159E_Papillon}
\end{figure}

\begin{figure}
 \begin{center}
  \includegraphics[width=9cm]{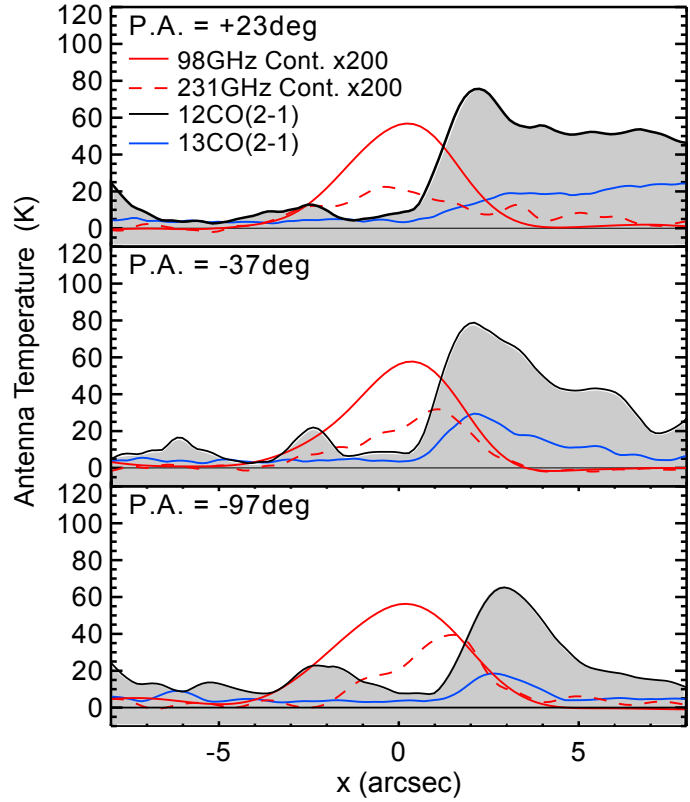}
 \end{center}
 \caption{ 
Antenna temperature distributions along three different position angles around Papillon HII region (see dashed boxs in Figure \ref{N159E_Papillon}b). 
Offset center is set the central position of Papillon Nebula. 
Red solid and dash lines show 200 times of the antenna temperature of 98~GHz and 231~GHz continuum emission. 
Black and blue lines show the peak antenna temperature distribution of $^{12}$CO(2-1) and $^{13}$CO(2-1). }
\label{N159E_Papillon_Slice}
\end{figure}

\begin{figure}
 \begin{center}
  \includegraphics[width=17cm]{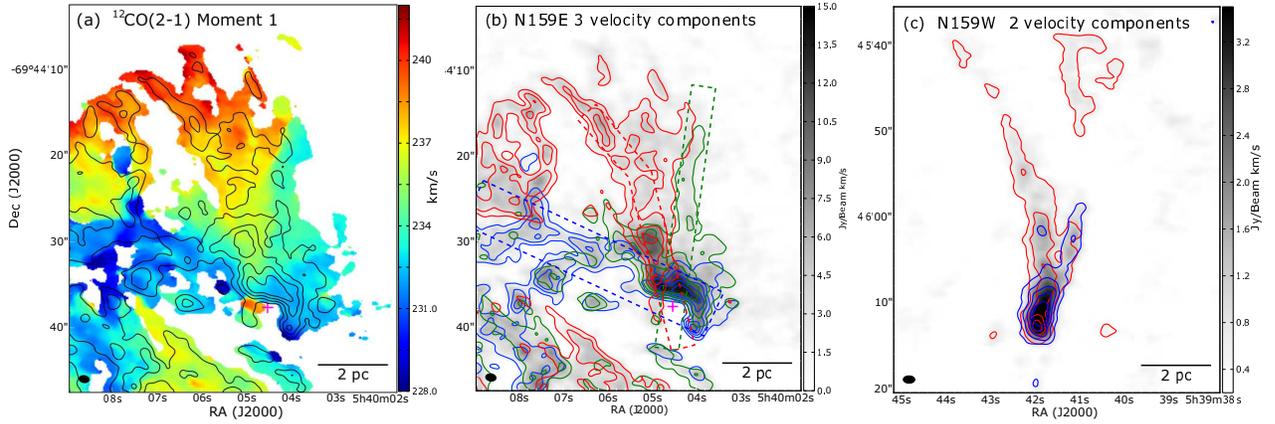}
 \end{center}
 \caption{(a) Moment 1 map of the 12CO (J=2-1). Black contour show the moment0 map with levels of 10, 20, 40, 60, 80sigma (1 sigma = 0.15Jy/Beam km s$^{-1}$). 
(b) Three velocity components of $^{12}$CO(J=2-1). 
Blue, green, and red contour show the integrated intensity map with -8.0~km~s$^{-1}$ - -1.6~km~s$^{-1}$, -  -1.~4km~s$^{-1}$ - 1.0~km~s$^{-1}$, and 1.2~km~s$^{-1}$ - 7.8~km~s$^{-1}$, respectively.  
(c) Two velocity components of $^{13}$CO(J=2-1) in N159W  }
\label{N159E_Papillon_Filaments}
\end{figure}

\begin{figure}
 \begin{center}
  \includegraphics[width=17cm]{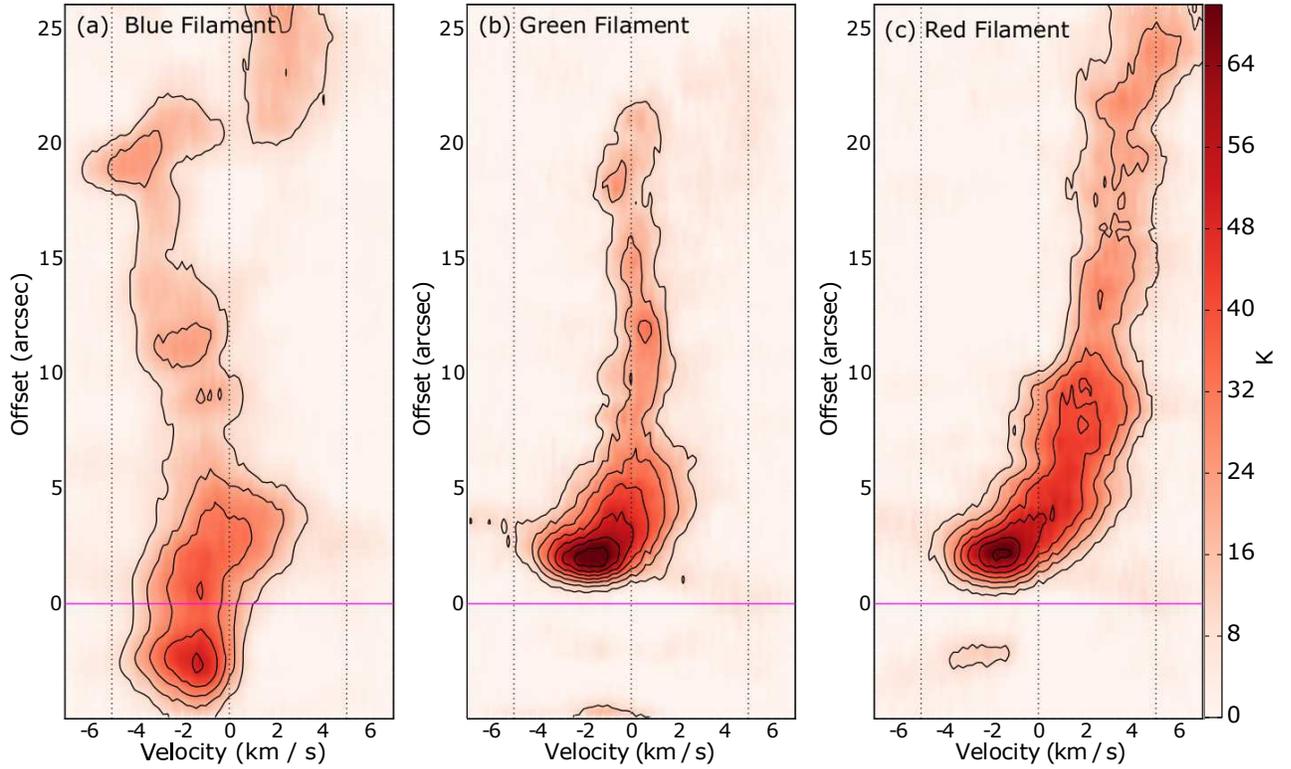}
 \end{center}
 \caption{
Position-Velocity diagrams of blue, green, and red filaments of N159E (magenta rectangles in figure \ref{N159E_Papillon_Filaments}b).
Widths of the region of blue, green and red filaments are 0.7pc, 0.7pc and 1.4pc, respectively.   
The offset positions of Papillon (magenta cross in figure 5b) show magenta horizontal lines (offset = 0 arcsec). 
}
\label{N159E_Papillon_PV}
\end{figure}

\begin{enumerate}

\item The prominent feature of the molecular gas distribution in N159E is three belt-like structure extending from the northeast to the southwest with length of 10~pc -- 20~pc (see Figure \ref{N159E_mom0}). 
It is well coincident with the dark lane in the optical H$\alpha$ images (see Figure \ref{N159_Halpha}).  
The total molecular mass of N159E in the present map is calculated to be $0.92 \times 10^{5}$~M$_{\odot}$ from the $^{13}$CO(2-1) intensity.

\item N159E is the complex of filamentary clouds with the width of $\sim$1~pc scale and projection length of several pc scale (see Figure \ref{N159E_channel}). 
Typical molecular mass and velocity dispersion of the filamentary clouds are a few thousand M$_{\odot}$ and $\sim$1~km s$^{-1}$. 

\item We found a molecular hole with the radius of $\sim$ 0.5~pc around the $\sim$40~M$_{\odot}$ massive YSO in N159E (so-call Papillon Nebula YSO). 
This molecular hole is filled by 98~GHz free-free emission and H30$\alpha$ hydrogen recombination emission. 
Temperature of molecular gas around the hole (i.e. interface) is as high as $\sim$ 80~K. 
This massive YSO has enough luminosity to ionize the parental filamentary clouds within the life time. 
It indicates that Papillon Nebula YSO has just moved to the initial phase of dispersal of the parental molecular cloud from the protostellar phase. 

\item Papillon Nebula YSO is clearly located at the overlapping point of three filamentary clouds of diffrent celocity. 
This kinematic structure suggests that formation of the massive protostar was induced by the collision of filamentary clouds. 
Similar kinematic structure was reported by \citep{Fukui2015} in N159W-S. 
These two objects are only massive YSO dandidates that have the mass larger than 35~M$_{\odot}$ in N159E and N159W region. 
This fact suggests strongly that such collision of filamentary clouds is primary or important mechanism of very high-mass star formation.

\item Collision of filamentary clouds is expected to occur frequently during the GMC evolutionary timescale of 10~Myrs if we assume relative speed 10~km s$^{-1}$ and length of 10~pc. 
It suggests that high-mass star formation induced by collision of filamentary cloud occur generally during the life of the molecular cloud.


\end{enumerate}

The authors thank the anonymous referee for his/her helpful comments. 
This paper makes use of the following ALMA data: ADS/JAO.ALMA \#2012.1.00554.S. ALMA is a partnership of ESO (representing its member states), NSF (USA) and NINS (Japan), together with NRC (Canada), NSC and ASIAA (Taiwan), and KASI (Republic of Korea), in cooperation with the Republic of Chile. 
The Joint ALMA Observatory is operated by ESO, AUI/NRAO and NAOJ. This work was supported by JSPS KAKENHI grant numbers 15555041, 22244014, 22540250, 22740127, 23403001, 24224005, 26247026; by JSPS and by the Mitsubishi Foundation. 

\newpage

\bibliographystyle{mn2e}
\bibliography{master}

\begin{deluxetable}{lllrrr}
\rotate
\tablecolumns{6}
\tablewidth{0pc}
\tablecaption{Properties of YSOs in N159E \label{t1}}
\tablehead{
\colhead{IRAC Name} & \colhead{Other Name}   & \colhead{$\mathrm{M_{*,avg}}$ [$\mathrm{M_{\odot}}$]}    & \colhead{$\mathrm{L_{*,avg}}$ [$\mathrm{L_{\odot}}$]} &
\colhead{$\mathrm{t_{best}}$ [years]}    & \colhead{Stage}}
\startdata
J054004.40-694437.6 & Papillon Nebula YSO & 35 $\pm$ 2    & (1.8 $\pm$ 0.1) $\times 10^{5}$ &  $4.3 \times 10^{5}$ & I \\ 
J054009.49-694453.5 &                     & 18 $\pm$ 2    & (4.5 $\pm$ 0.6) $\times 10^{4}$ &  $2.0 \times 10^{4}$ & I \\ 
J054000.68-694439.2 &                 & 11.2          & 4.7 $\times 10^{4}$             &                     & I \\ 
J054005.26-694400.3 &                 & 9.5 $\pm$ 0.2 & (5.4 $\pm$ 0.5) $\times 10^{3}$ &                     & I \\ 
\enddata
\end{deluxetable}

\begin{deluxetable}{lcccccc}
\rotate
\tablecolumns{7}
\tablewidth{0pc}
\tablecaption{Properties Derived from H30$\alpha$ \label{t2}}
\tablehead{
\colhead{Object} & \colhead{$\mathrm{n_{e}}$ [$\mathrm{cm^{-3}}$]}   & \colhead{$\mathrm{M_{ionized}}$ [$\mathrm{M_{\odot}}$]} &
\colhead{EM [$\mathrm{pc\;cm^{-6}}$]} &  \colhead{U [$\mathrm{pc\;cm^{-2}}$]} &
\colhead{$\mathrm{N_{continuum}}$ [$\mathrm{s^{-1}}$]} & \colhead{Type}} 
\startdata
Papillon Nebula YSO & 2.2 $\times 10^{3}$ & 65 & 3.9 $\times 10^{6}$ & 153 & $\ge$1.1 $\times 10^{50}$ & $\ge$O3 \\
N159W-N YSO (or YSO-N) & 2.6 $\times 10^{3}$ & 154 & 6.3 $\times 10^{6}$  & 168 & $\ge$1.5 $\times 10^{50}$ & $\ge$O3 \\ 
\enddata
\end{deluxetable}

\begin{deluxetable}{lllrrc}
\rotate
\tablecolumns{6}
\tablewidth{0pc}
\tablecaption{Properties of YSO in N159W \label{t3}}
\tablehead{
\colhead{IRAC Name} & \colhead{Other Name} &
\colhead{$\mathrm{M_{*,avg}}$ [$\mathrm{M_{\odot}}$]} &
\colhead{$\mathrm{L_{*,avg}}$ [$\mathrm{L_{\odot}}$]} &
\colhead{$\mathrm{t_{best}}$ [years]} &
\colhead{Stage } }
\startdata
J053937.56-694525.4  & N159W-N YSO (or YSO-N)& 31 $\pm$ 8     & (1.4 $\pm$ 0.4) $\times 10^{5}$  & $5.9 \times 10^{4}$ & I \\ 
J053941.89-694612.0  & N159W-S YSO (or YSO-S)& 37 $\pm$ 2     & (2.0 $\pm$ 0.3) $\times 10^{5}$  & $3.9 \times 10^{4}$ & I \\ 
J053943.75-694540.2  &         & 9.2 $\pm$ 1.3  & (5.3 $\pm$ 2.3) $\times 10^{3}$  &                    & I \\ 
J053940.48-694517.3  &         & 11.6 $\pm$ 1.0 & (4.2 $\pm$ 1.1) $\times 10^{3}$  &                    & I \\ 
J053940.78-694632.0  &         & 9.4 $\pm$ 2.7  & (5.7 $\pm$ 2.8) $\times 10^{3}$  &                    & II \\ 
J053935.99-694604.1  &         & 18.5 $\pm$ 1.4 & (4.5 $\pm$ 0.04) $\times 10^{4}$ &                    & I \\ 
J053937.53-694609.8  &         & 30 $\pm$ 4     & (1.3 $\pm$ 0.4) $\times 10^{5}$  & $2.6 \times 10^{5}$ & I \\
J053937.04-694536.7  &         & 28             &  1.0             $\times 10^{5}$ & $6.1 \times 10^{4}$ & I \\
\enddata
\end{deluxetable}

\newpage

\end{document}